\documentclass[%
 aip,
 amsmath,
 amssymb,
 reprint,%
]{revtex4-1}

\usepackage{graphicx}

\usepackage{float}
\usepackage{subfigure}
\usepackage[colorlinks,breaklinks,linkcolor=blue,anchorcolor=blue,citecolor=blue,urlcolor=blue]{hyperref}
\usepackage{dcolumn}
\usepackage{bm}

\usepackage[utf8]{inputenc}
\usepackage[T1]{fontenc}
\usepackage{mathptmx}
\usepackage{etoolbox}
\usepackage{tikz}
\usetikzlibrary{shapes}

\makeatother
\begin{document}

\preprint{AIP/123-QED}

\title[Simulation of chiral motion of excitation within the ground-state manifolds of neutral atoms]{Simulation of chiral motion of excitation within the ground-state manifolds of neutral atoms}

\author{Hao-Yuan Tang}
\affiliation{Center for Quantum Sciences and School of Physics, Northeast Normal University, Changchun, 130024, China}

\author{Xiao-Xuan Li}
\affiliation{Sino-European Institute of Aviation Engineering, Civil Aviation University of China, Tianjin 300300, China}

\author{Jia-Bin You}
\email[]{Authors to whom correspondence should be addressed: you\_jiabin@ihpc.a-star.edu.sg and shaoxq644@nenu.edu.cn}
\affiliation{Institute of High Performance Computing, A*STAR (Agency for Science, Technology and Research), 1 Fusionopolis Way, Connexis, Singapore 138632}

\author{Xiao-Qiang Shao}
\email[]{Authors to whom correspondence should be addressed: you\_jiabin@ihpc.a-star.edu.sg and shaoxq644@nenu.edu.cn}
\affiliation{Center for Quantum Sciences and School of Physics, Northeast Normal University, Changchun, 130024, China}
 \affiliation{Center for Advanced Optoelectronic Functional Materials Research, and Key Laboratory for UV Light-Emitting Materials and Technology of Ministry of Education, Northeast Normal University, Changchun 130024, China}

\date{\today}

\begin{abstract}
Laser-induced gauge fields in neutral atoms serve as a means of mimicking the effects of a magnetic field, providing researchers with a platform to explore behaviors analogous to those observed in condensed matter systems under real magnetic fields. Here, we propose a method to generate chiral motion in atomic excitations within the neutral atomic ground-state manifolds. This is achieved through the application of polychromatic driving fields coupled to the ground-Rydberg transition, along with unconventional Rydberg pumping. The scheme offers the advantage of arbitrary adjustment of the effective magnetic flux by setting the relative phases between different external laser fields. Additionally, the effective interaction strength between the atomic ground states can be maintained at 10 kHz, surpassing the capabilities of the previous approach utilizing Floquet modulation. Notably, the proposed method can be readily extended to implement a hexagonal neutral atom lattice, serving as the fundamental unit in realizing the Haldane model.
\end{abstract}

\maketitle

%

\section{Introduction}
Quantum simulation refers to the use of a controllable quantum system to model and understand the behavior of another quantum system that may be difficult to study directly.\cite{nature24622,nature24654,nature41586,Randall20211474,PhysRevA.91.022311,Argüello-Luengo2019215,PhysRevLett.99.030603,PhysRevLett.125.120605,PhysRevLett.107.260501} A key objective of this approach is to explore exotic phases arising from nontrivial topology in quantum lattice models, such as topological insulators (TIs).\cite{PhysRevLett.61.2015,RevModPhys.82.3045,Jotzu2014237,Wang:15} These insulators showcase distinctive phases of matter, including the quantum Hall effect (QHE),\cite{PhysRevLett.45.494,PhysRevLett.48.1559} and offer insight into the intrinsic properties of band structures. This methodology facilitates the integration of gauge fields into diverse physical systems, \cite{PhysRevLett.107.255301,Kolovsky_2011,PhysRevX.4.031027,Schmidt:15} enabling the induction of chiral motion in atomic excitations. In earlier work, the transfer of chiral Fock states in the QED architecture of circuits was achieved.\cite{PhysRevA.82.043811} Subsequently, superconducting qubits \cite{Roushan2017146,wang2019synthesis,Liu2020,PhysRevA.102.032610,PhysRevA.106.053714} and ultracold atom-based platforms \cite{RevModPhys.83.1523,AIDELSBURGER2018394,RevModPhys.91.015005} became the widely used simulation platforms in the artificial gauge field. Recently, cavity-magnonics system~\cite{PhysRevA.106.033711} and chiral quantum spin liquids \cite{bauer2014chiral,PhysRevLett.124.217203,PhysRevB.108.075118} had also been proposed to study these issues.

Compared with the above systems, neutral-atom systems with Rydberg interactions are considered a promising choice.\cite{RevModPhys.82.2313,Saffman_2016,barredo2016atom,wu2021concise,morgado2021quantum,Shi_2022,shao2024rydberg} The strong and controllable interactions of Rydberg atoms offer significant advantages in simulating various many-body problems, particularly in exploring exotic phases and dynamics. For example, topological bands and Chern insulators have been implemented in Rydberg atoms,~\cite{PhysRevA.91.053617,Weber_2018,PRXQuantum.3.030302,PhysRevA.108.053107} the Peierls phase has been realized in Rydberg systems with exchange interactions,~\cite{PhysRevA.97.033414,PhysRevX.10.021031} and gauge fields have been synthesized in Rydberg-atom arrays.~\cite{Ohler_2022,PhysRevResearch.4.L032046,PhysRevA.109.032622} 
We note that the schemes mentioned above, utilizing neutral atoms for simulation, can encode either the qubit into a dipole-dipole coupled Rydberg states or the ground state and Rydberg state of the atom. However, it is essential to acknowledge that Rydberg states have a finite lifetime because of spontaneous emission, leading to decay into low-lying excited states, and blackbody radiation transfers the atom to nearby Rydberg states. Additionally, electrons in the Rydberg state are more susceptible to external fields, introducing additional sources of decoherence.

Chiral excitation current has been achieved in different systems, exhibiting unique physical properties.\cite{Roushan2017146,Liu2020,wang2023chiral,wang2024ground} Recently, some of us have proposed a scheme for quantum state transfer within the ground-state manifolds of neutral atoms.~\cite{PhysRevA.105.032417} This scheme proves effective in simulating various single-body physics phenomena while simultaneously minimizing the impact of the Rydberg excited state. In an equilateral triangular structure, a chiral motion of excitation becomes observable through the periodic modulation of weak pulses. However, the scalability of this method for achieving quantum simulations with a larger number of atoms is limited. Furthermore, there is no guarantee of consistent periods for two chiral evolutions. In this work, we address this limitation by driving the atoms through the application of polychromatic fields,\cite{PhysRevResearch.4.L032046} combined with the unconventional Rydberg pumping. \cite{PhysRevA.98.062338,PhysRevA.102.053118,PhysRevApplied.20.014014,10.1063/5.0192602} The induction of an effective magnetic flux arises from complex-valued hopping amplitudes between different atoms on the sites, leading to the Peierls phase. Consequently, we can adjust the Peierls phase arbitrarily by setting the relative phases between different external laser fields. This approach allows us to generate a gauge field in an equilateral triangle system composed of three atoms, thereby breaking the time-reversal symmetry (TRS). Compared to the approach in Ref.~\onlinecite{PhysRevA.105.032417}, we eliminate the need to periodically switch on and off weak driving fields, enhancing the scalability of the present scheme.

\section{Realization of Chiral Motion in Three Atoms}
\subsection{Model}

\begin{figure}
    \centering
    \includegraphics[width=0.95\linewidth]{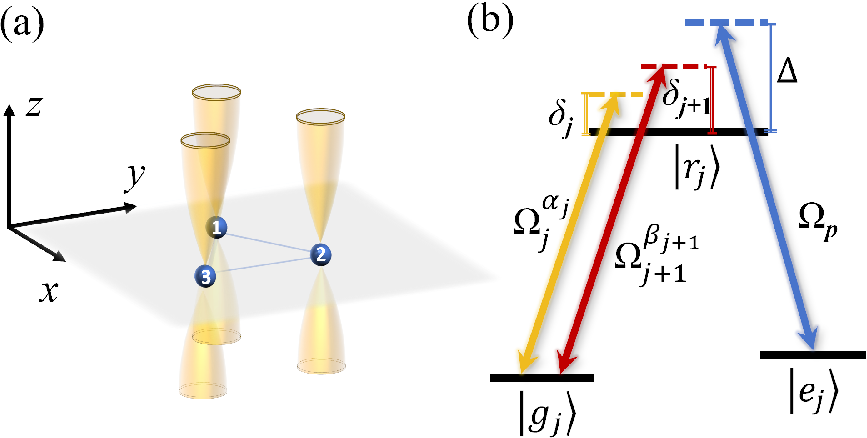}
    \caption{(a) Three neutral atoms are arranged in an equilateral triangle in the $x-y$ plane. (b) The atomic structure in our scheme has three levels: $|g_j\rangle$, $|e_j\rangle$, and $|r_j\rangle$. Each atom is driven by one global laser field with real Rabi frequency $\Omega _p$, detuned by $\Delta$, and two local laser fields with complex Rabi frequencies $\Omega _j^{\alpha_j}$ and $\Omega _{j+1}^{\beta_{j+1}}$, detuned by $\delta_j$ and $\delta_{j+1}$, respectively. }
\label{fig1}
\end{figure}

The configuration of our setup is illustrated in Fig.~\ref{fig1}(a), where three $^{87}$Rb atoms form an equilateral triangle structure in the $x-y$ plane, maintaining uniform spacing $R$ between each pair of atoms, and the quantization axis is $z$.  In Fig.~\ref{fig1}(b), the specific configuration of each atom is described, highlighting the ground states $|g\rangle=| 5S_{1/2},\ F=1, m_F=0 \rangle$ and $|e\rangle=| 5S_{1/2},\ F=2,\ m_F=0 \rangle $, in addition to the Rydberg state $|r\rangle=| 60S_{1/2}, \ m_J=1/2 \rangle $. The coupling between the ground state $|g\rangle$ of the $j$-th atom and the Rydberg state $|r\rangle$ involves two off-resonant local laser fields along the axis $z$ which are characterized by complex Rabi frequencies $\Omega^{\alpha_j} _j=\Omega _je^{i\alpha_j}$ and $\Omega^{\beta_{j+1}} _{j+1}= \Omega _{j+1}e^{i\beta_{j+1}}$ with positive real $\Omega_j$ $(j=1,2,3)$, and the corresponding detunings are denoted by $\delta _j$ and $\delta _{j+1}$, with the condition $\Omega^{\beta_4} _{4} =\Omega _{1}e^{i\beta_1}$ and $\delta _{4}=\delta _{1}$ for completeness. Additionally, the state $|e\rangle$ is coupled to $|r\rangle$ through a global laser field along the axis $z$, which is characterized by a real Rabi frequency $\Omega _p$ and is detuned by $\Delta$. The van der Waals (vdW) interaction between Rydberg states takes the form of $\mathcal{U}_{i,j}=-C_6/R^6$, where the dispersion coefficient $C_6/(2\pi)\approx-135~{\rm GHz}\cdot\mu {\rm m}^6$ (${\hbar }=1$), as given by the second-order non-degenerate perturbation theory, \cite{SIBALIC2017319} and $R$ is the interatomic distance.  Assuming $R\approx2.96$ $\mu {\rm m}$, we have $\mathcal{U}_{j,j+1}/(2\pi)=200$ MHz.
Now, the full Hamiltonian of the proposed model reads
\begin{eqnarray}\label{Hamiltonian_1}
H_I&=&\sum_{j=1}^3[(\Omega _j^{\alpha_j}e^{-i\delta _jt}+\Omega _{j+1}^{\beta_{j+1}}e^{-i\delta _{j+1}t})| r_j \rangle\langle g_j|\nonumber\\&&
+\Omega _p| r_j \rangle\langle e_j|e^{-i\Delta t}+{\rm H.c.}]+\frac{1}{2}\sum_{i\neq j}\mathcal{U}_{i,j} |r_ir_{j} \rangle\langle r_ir_{j}|.
\end{eqnarray}
For simplicity, we assume that the Rydberg state directly decays into $|g\rangle$ and $|e\rangle$, with equal branching ratios for the spontaneous emission rates. This assumption establishes the subsequent consideration of independent decay channels
$L_1=\sqrt{\gamma/2}|g_1\rangle\langle r_1|$, $L_2=\sqrt{\gamma/2}|e_1\rangle\langle r_1|$, $L_3=\sqrt{\gamma/2}|g_2\rangle\langle r_2|$, $L_4=\sqrt{\gamma/2}|e_2\rangle\langle r_2|$, $L_5=\sqrt{\gamma/2}|g_3\rangle\langle r_3|$, and $L_6=\sqrt{\gamma/2}|e_3\rangle\langle r_3|$. Therefore, the system's evolution is described by the Markovian master equation:
\begin{equation}\label{master_1}
\dot\rho =-i\left[ H_I,\rho \right]+\sum_{k=1}^6L_k\rho L_k^{\dag}-\frac{1}{2}\{ L_k^{\dag}L_k,\rho \} .
\end{equation}
Now, we designate the state $|e\rangle$ in the ground-state manifold of neutral atoms as an effective ``excited'' state. Subsequently, we will elucidate the derivation process for the effective chiral dynamics of the dominant system within the single-``excitation'' subspace $\{|egg\rangle, |geg\rangle, |gge\rangle\}$.
When considering a rotating frame
with respect to $U=\exp[-i\sum_{i\neq j}\mathcal{U}_{i,j}/2|r_ir_{j} \rangle\langle r_ir_{j}|t]$, the high-frequency oscillation term can be disregarded due to the limit of large detuning $\mathcal{U}_{i,j}=\Delta \gg \{ \Omega _p, \Omega _j, \delta _j \}$. As a result, the Hamiltonian of Eq.~(\ref{Hamiltonian_1}) can be reformulated as (see Appendix \ref{Ap3} for details):
\begin{eqnarray}\label{Hamiltonian_2}
H'_{I}&=&\sum_{j=1}^3  \Omega _{j}^{\alpha_j}{\sigma}_{j}^{rg}(P_{j+1}^{e}P_{j+2}^{g}+P_{j+1}^{g}P_{j+2}^{e})e^{-i\delta_jt}\nonumber\\&&
+\Omega _{j+1}^{\beta_{j+1}}{\sigma}_{j}^{rg}(P_{j+1}^{e}P_{j+2}^{g}+P_{j+1}^{g}P_{j+2}^{e})e^{-i\delta_{j+1}t} \nonumber\\&&
+\Omega_p \sigma _{j}^{re}(P_{j+1}^{g}P_{j+2}^{r}+P_{j+1}^{r}P_{j+2}^{g})+{\rm H.c.} ,
\end{eqnarray}
with projection operator $P_j^k=|k_j\rangle\langle k_j|$ and transition operator $\sigma _j^{ab}=|a_j\rangle\langle b_j|$ (here $P_j^k=P_{j-3}^k$ and $\sigma _j^{ab}=\sigma _{j-3}^{ab}$ for $j>3$). Eq.~(\ref{Hamiltonian_2}) can be transformed into a time-independent form by moving to another rotating frame defined by $U'=\exp\{ -i[\delta_{1}(|egr \rangle\langle egr|+| rge \rangle \langle rge |+| rgr \rangle \langle rgr | )t+\delta_{2}(| erg \rangle\langle erg |+| reg \rangle \langle reg |+| rrg \rangle \langle rrg | )t+\delta_{3}(| ger \rangle\langle ger |+| gre \rangle \langle gre |+| grr \rangle \langle grr | )t ] \}$. 
The resulting Hamiltonian is expressed as $H''_{I}=H_0+V_++V_+^{\dag}$, where 
\begin{eqnarray}
H_0&=&\Omega _p(| rgr \rangle\langle egr |+| rrg \rangle\langle erg |+| grr \rangle\langle ger |+| rrg \rangle\langle reg |\nonumber\\&&+| rgr \rangle\langle  rge |+| grr \rangle\langle gre | ) +{\rm H.c.} \nonumber\\&&
-\delta_{1}(|egr \rangle\langle egr|+| rge \rangle \langle rge |+| rgr \rangle \langle rgr | )\nonumber\\&&
-\delta_{2}(| erg \rangle\langle erg |+| reg \rangle \langle reg |+| rrg \rangle \langle rrg | )\nonumber\\&&
-\delta_{3}(| ger \rangle\langle ger |+| gre \rangle \langle gre |+| grr \rangle \langle grr | ),
\end{eqnarray}
which is essentially a block diagonalization matrix defined in three different subspaces  $\{|egr\rangle, |rgr\rangle, |rge\rangle\}$, $\{|erg\rangle, |rrg\rangle, |reg\rangle\}$, and $\{|gre\rangle, |grr\rangle, |ger\rangle\}$,
and $V_++V_+^{\dag}$ can be regarded as a probe field for coupling the atomic ground state to the single-excited Rydberg state, taking the form of
\begin{eqnarray}
V_+&=&\Omega _1^{\alpha_1} |rge \rangle\langle gge| +\Omega _2^{\beta_2} | reg \rangle\langle geg | +\Omega _2^{\alpha_2}| erg \rangle\langle egg |\nonumber\\&& +\Omega _3^{\beta_3} | gre \rangle\langle gge |+\Omega _3^{\alpha_3}| ger \rangle\langle geg |+\Omega _1^{\beta_1}| egr \rangle\langle egg |.\nonumber\\&&
\end{eqnarray}

Assuming $\{\Omega_p, |\delta_j|\}\gg \Omega_j$ and employing the effective operator formalism for open quantum systems, \cite{PhysRevLett.106.090502,PhysRevA.85.032111,PhysRevA.88.032317} the effective Hamiltonian and the corresponding decay operators can be expressed by $H_{\textrm{eff}}=-\frac{1}{2}[ V_+^{\dag}H_{0}^{-1}V_++V_+^{\dag}( H_{0}^{-1} ) ^{\dag}V_+ ]$ and $L_{\textrm{eff}}^k=L_k H_{NH}^{-1} V_+$, where $H_{NH}=H_0-\frac{i}{2}\sum_j{L_j^{\dag}L_j}$. Within the subspace of consideration, the concrete form of the effective coherent part reads
\begin{equation}\label{Hamiltonian_4}
H_{\text{eff}}=\sum_{j=1}^3{J_{j,j+1} \sigma _{j}^{-}\sigma _{j+1}^{+}+{\rm H.c.}} ,
\end{equation}
where 
\begin{equation}\label{7}
J_{j,j+1}=\frac{\Omega _{j+1}^{\alpha_{j+1}}(\Omega _{j+1}^{\beta_{j+1}})^{\ast}\Omega _p^2}{\delta _{j+1}^{3}-2\delta _{j+1}\Omega _p^2},
\end{equation}
which describes the effective coupling between ground-state atoms that carries a phase factor $e^{i\phi_{j,j+1}}=e^{i(\alpha_{j+1}-\beta_{j+1})}$ and the pseudo spin raising operator $\sigma _j^+=| e_j \rangle\langle g_j|$. Note that in Eq.~(\ref{Hamiltonian_4}), the Stark-shift terms $[\Omega _1^2( \delta _{1}^{2}-\Omega _{p}^{2} )/( \delta _{1}^{3}-2\delta _1\Omega _{p}^{2} )+\Omega _2^2( \delta _{2}^{2}-\Omega _{p}^{2} )/( \delta _{2}^{3}-2\delta _2\Omega _{p}^{2} )] | egg \rangle\langle egg |$, $[\Omega _2^2( \delta _{2}^{2}-\Omega _{p}^{2} )/( \delta _{2}^{3}-2\delta _2\Omega _{p}^{2} )+\delta _{3}^{2}-\Omega _{p}^{2} )/( \delta _{3}^{3}-2\delta _3\Omega _{p}^{2}]| geg \rangle\langle geg |$, and $[\Omega _1^2( \delta _{1}^{2}-\Omega _{p}^{2} )/( \delta _{1}^{3}-2\delta _1\Omega _{p}^{2} )+\Omega _3^2( \delta _{3}^{2}-\Omega _{p}^{2} )/( \delta _{3}^{3}-2\delta _3\Omega _{p}^{2} )] | gge \rangle\langle gge |$ have been neglected since they can be easily canceled by introducing additional lasers.
Correspondingly, the effective master equation is given by
\begin{equation}\label{master_2}
\dot\rho =-i \left[ H_{\textrm{eff}}, \rho \right]+\sum_{k=1}^6{L_{\textrm{eff}}^{k}\rho L_{\textrm{eff}}^{k\dag}-\frac{1}{2}\{ L_{\textrm{eff}}^{k\dag}L_{\textrm{eff}}^{k},\rho \}},
\end{equation}
where the jump operators describing the dissipative part of the dynamics are
$L_{\textrm{eff}}^{1}=L_{1,1}^{(2)}+L_{3,3}^{(1)}$, $L_{\textrm{eff}}^{2}=L_{2,1}^{(2)}+L_{4,3}^{(1)}$, $L_{\textrm{eff}}^{3}=L_{3,1}^{(2)}+L_{1,2}^{(3)}$, $L_{\textrm{eff}}^{4}=L_{4,1}^{(2)}+L_{2,2}^{(3)}$, $L_{\textrm{eff}}^{5}=L_{3,2}^{(3)}+L_{1,3}^{(1)}$, and $L_{\textrm{eff}}^{6}=L_{4,2}^{(3)}+L_{2,3}^{(1)}$, 
where
\begin{eqnarray}
L_{1,j}^{(j)}&=&\Gamma _1^{(j)}| g_je_{j+1} \rangle\langle g_je_{j+1} |+\Gamma _2^{(j)}| g_je_{j+1} \rangle\langle e_jg_{j+1} |\nonumber\\&&
+\Gamma _3^{(j)}(| g_jr_{j+1} \rangle\langle g_je_{j+1} |+| g_jr_{j+1} \rangle\langle e_jg_{j+1} |),
\end{eqnarray}
\begin{eqnarray}
L_{2,j}^{(j)}&=&\Gamma _1^{(j)}| e_je_{j+1} \rangle\langle g_je_{j+1} |+\Gamma _2^{(j)}| e_je_{j+1} \rangle\langle e_jg_{j+1} |\nonumber\\&&
+\Gamma _3^{(j)}(| e_jr_{j+1} \rangle\langle g_je_{j+1} |+| e_jr_{j+1} \rangle\langle e_jg_{j+1} |),
\end{eqnarray}
\begin{eqnarray}
L_{3,j}^{(j)}&=&\Gamma _1^{(j)}| e_jg_{j+1} \rangle\langle e_jg_{j+1} |+\Gamma _2^{(j)}| e_jg_{j+1} \rangle\langle g_je_{j+1} | \nonumber\\&&
+\Gamma _3^{(j)}(| r_jg_{j+1} \rangle\langle g_je_{j+1} |+| r_jg_{j+1} \rangle\langle e_jg_{j+1} |),
\end{eqnarray}
\begin{eqnarray}
L_{4,j}^{(j)}&=&\Gamma _1^{(j)}| e_je_{j+1} \rangle\langle e_jg_{j+1} |+\Gamma _2^{(j)}| e_je_{j+1} \rangle\langle g_je_{j+1} |\nonumber\\&&
+\Gamma _3^{(j)}(| r_je_{j+1} \rangle\langle g_je_{j+1} |+| r_je_{j+1} \rangle\langle e_jg_{j+1} |),
\end{eqnarray}
with the effective spontaneous emission rates $\Gamma _1^{(j)}=|\Omega_j\chi ( \gamma ^2-3i\gamma \delta_j -2\delta_{j}^{2}+2\Omega _{p}^{2} ) /( \gamma -2i\delta_j )| $, $\Gamma _2^{(j)}=|2\Omega _j\Omega _p^{2}\chi/( \gamma -2i\delta_j)|$, and $\Gamma _3^{(j)}=|(\Omega _j^{\alpha_j}+\Omega _j^{\beta_j})\Omega _p\chi/2 |$, where $\chi =\sqrt{2\gamma}/( \gamma ^2-3i\gamma \delta _j-2\delta _{j}^{2}+4\Omega _{p}^{2} )$.

\subsection{Numerical simulations}
\begin{figure}
    \centering
    \includegraphics[width=1\linewidth]{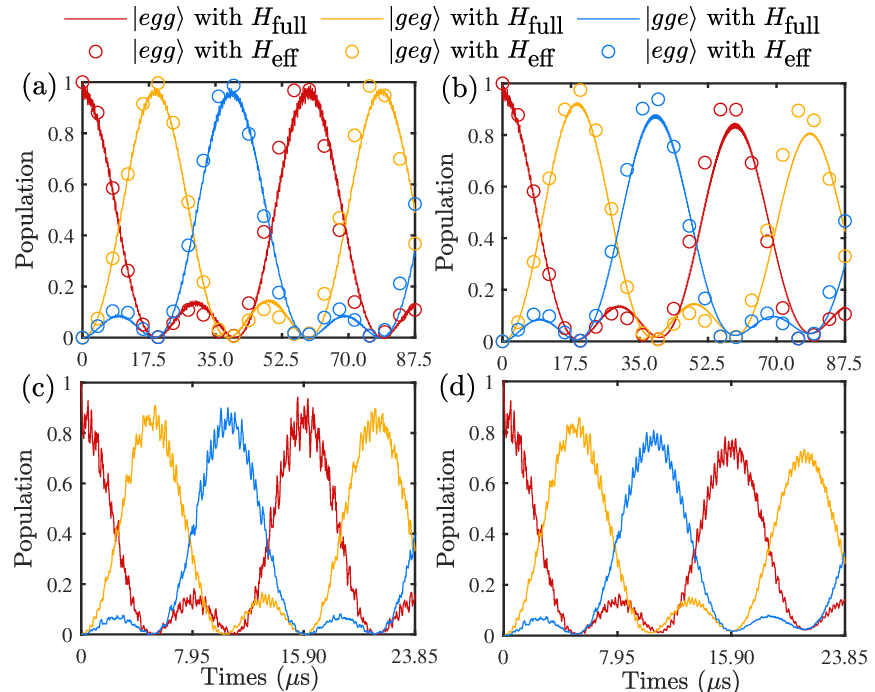}
    \caption{The chiral motion of atomic excitation among the three atoms. (a) and (b) show chiral motion in the clockwise direction of the atoms, and the solid line represents the simulation carried out using the full master equation of Eq.~(\ref{master_1}), and the scatter denotes the simulation based on the effective master equation of Eq.~(\ref{master_2}). The parameters are $(\Omega _1,\Omega _2,\Omega _3,\Omega _p)/(2\pi)=(0.2,0.2,0.4,4)$ MHz, $(\alpha_1,\alpha_{j\ne 1},\beta_j)=(\pi/{2},0,0)$, $( \delta _1,\delta _2,\delta_3,\Delta)/ (2\pi)=(-4,4,8,200)$ MHz, $\gamma/(2\pi)=6.92$ kHz. (c) and (d) show population oscillations in a clockwise direction as governed by the master equation without and considering atomic decay. The parameters are changed to $(\Omega _1,\Omega _2,\Omega _3,\Omega _p)/(2\pi)=(0.4,0.4,0.8,4)$ MHz, $(\alpha_1,\alpha_{j\ne 1},\beta_j)=(\pi/{2},0,0)$, $( \delta _1,\delta _2,\delta_3,\Delta)/ (2\pi)=(-4,4,8,200)$ MHz, $\gamma/(2\pi)=6.92$ kHz. }
\label{fig2}
\end{figure}
To ensure the generation of the ideal chiral motion of excitation, it is necessary to satisfy the following conditions: $|J_{12}|=|J_{23}|=|J_{31}|$ and synthetic flux $\Phi =(\alpha_2-\beta_2)+(\alpha_3-\beta_3)+(\alpha_1-\beta_1)=\pm \pi/2$. For $\Phi =\pi/{2}$, the atomic excitation $|e\rangle$ propagates in the clockwise direction $1\rightarrow 2\rightarrow 3\rightarrow 1$, and for $\Phi =-\pi/{2}$, the direction reverses, which breaks the TRS by the chiral motion of the system. In Figs.~\ref{fig2}(a) and~\ref{fig2}(b), we depict the population oscillations of states $|egg\rangle$, $|geg\rangle$, and $|gge\rangle$ versus the evolution time, starting from the state $|egg\rangle$. Under the parameters $\Omega _1=\Omega _2= 2\pi \times0.2$ MHz, $\Omega _3= 2\pi \times0.4$ MHz, $\Omega _p= 2\pi \times4$ MHz, $ \delta _1=-\delta _2= 2\pi \times4$ MHz, $ \delta _3= 2\pi \times8$ MHz, we have the hopping strengths between the single-``excitation'' states as $|J_{12}|=|J_{23}|=|J_{31}|=2\pi \times 10$ kHz, which far exceed the coupling strength $J_{\textrm{eff}}=2\pi\times 0.25~(0.50)$ kHz utilizing Floquet modulation.~\cite{PhysRevA.105.032417} We also choose $(\alpha_1,\alpha_{j\ne 1},\beta_j)=(\pi/{2},0,0)$ to ensure $\Phi =\pi/{2}$ to make the clockwise chiral motion of atomic excitation. The solid line represents the simulation carried out using the full master equation of Eq.~(\ref{master_1}), while the scatter denotes the simulation based on the effective master equation of Eq.~(\ref{master_2}). The aforementioned findings show that the evolution under the effective master equation is fundamentally consistent with the full master equation. This observation suggests that each atom can be accurately described within a two-level subspace in accordance with our proposed scheme.

In fact, we can further strengthen the flip-flop interaction between ground states by adjusting the parameters without striving for perfect matching between the effective master equation and the full master equation. To reduce the period of population oscillation of chiral excitation, we set the parameters to $\Omega _1=\Omega _2= 2\pi \times0.4$ MHz, $\Omega _3= 2\pi \times0.8$ MHz, $\Omega _p= 2\pi \times4$ MHz, $ \delta _1=-\delta _2= 2\pi \times4$ MHz, $ \delta _3= 2\pi \times8$ MHz, $\Delta=2\pi \times200$ MHz and $\gamma=2\pi \times6.92$ kHz. Figs.~\ref{fig2}(c) and \ref{fig2}(d) illustrate the clockwise chiral motion of atomic excitation under the parameters $(\alpha_1, \alpha_{j \ne 1}, \beta_j) = (\pi / 2, 0, 0)$, governed by the master equation without and with atomic decay, respectively. However, the system becomes more sensitive to spontaneous emission in this case due to the weaker suppression of Rydberg excitation. Therefore, although the effective hopping strength shown in Figs.~\ref{fig2}(a) and \ref{fig2}(b) is relatively small, it allows for a longer coherence time in the system.

As shown in Eq.~(\ref{7}), the synthetic flux can be adjusted by varying the phases $\alpha_j$ and $\beta_j$ of polychromatic driving fields, and the coupling strength depends solely on the Rabi frequency and detuning of the laser fields. Thus, this flexibility allows the scheme to be extended to a larger number of particles, allowing the simulation of more complex physical systems, such as the 2D triangular array.\cite{Ohler_2022} 

\section{Honeycomb lattice model\label{SecIII}}
\begin{figure}
    \centering
    \includegraphics[width=1\linewidth]{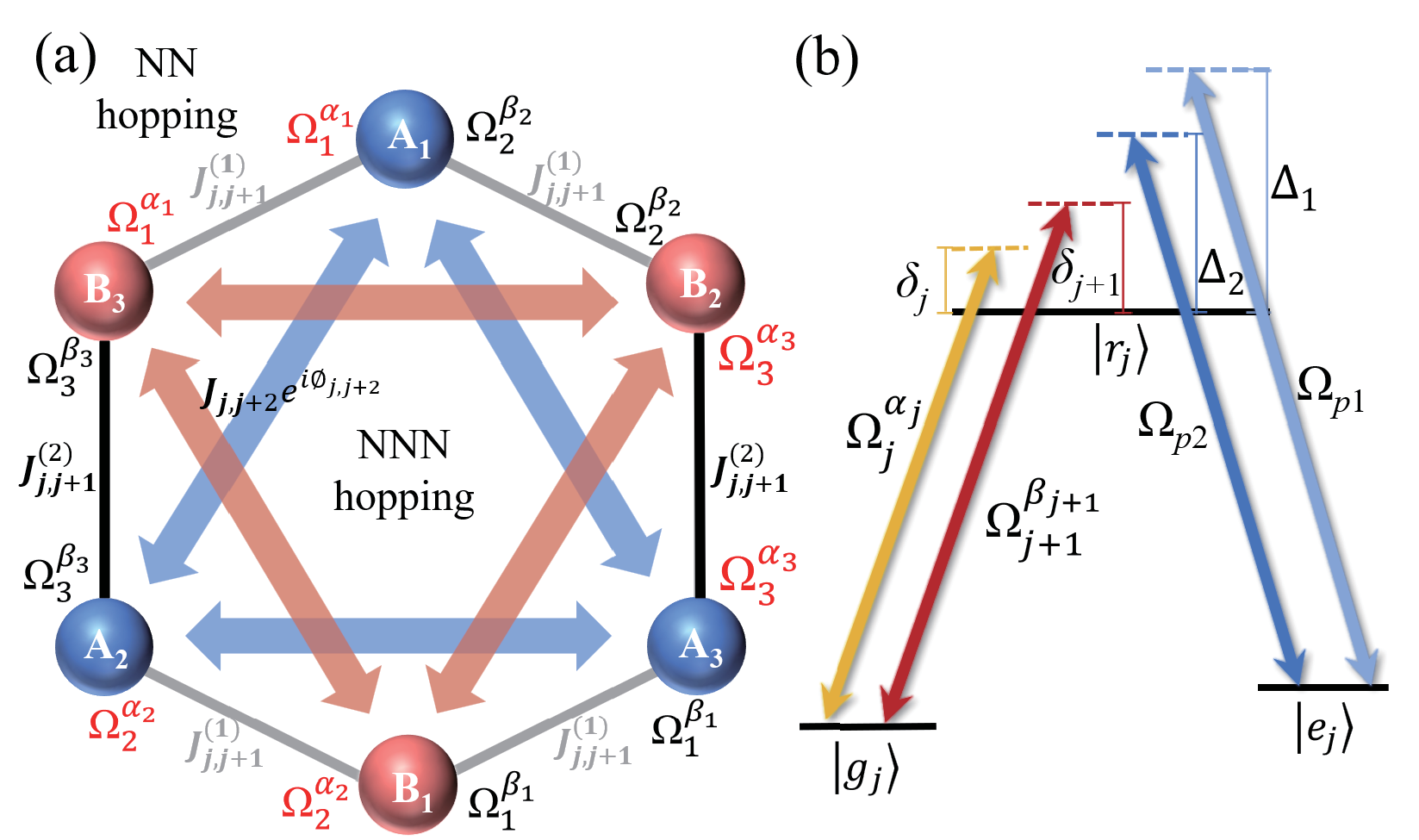}
    \caption{(a) The hexagonal neutral atom lattice, and the specific solution for applying local laser fields to achieve the basis unit of the Haldane model. (b) The laser-atom interaction models. }
\label{fig3}
\end{figure}

The QHE has led to the discovery of a new type of material, TIs. In 1988, the first TI model, the Haldane model, was proposed, \cite{PhysRevLett.61.2015} which revealed that the QHE can exhibit inherent properties of band structures rather than being caused by external magnetic fields. The Haldane model is a honeycomb tight-binding lattice based on breaking both inversion symmetry (IS) and TRS. The exotic phases that arise from this model are considered to be at the forefront of condensed matter physics. Within a single cell of the honeycomb lattice, the difference in energy between two sublattices may disrupt the IS, and the complex next-nearest-neighbor (NNN) tunnel coupling can lead to the disruption of TRS. Furthermore, the total synthetic magnetic flux amounts to zero, thereby preserving the translational symmetry of the lattice. 

The above-mentioned method for achieving chiral motion of excitation can be used directly to create a hexagonal neutral atom lattice, as shown in Fig.~\ref{fig3}(a), with two separate sublattices, A and B, which break the IS. The quantization axis is perpendicular to the plane of the atoms. The distance between nearest-neighbor (NN) atoms is denoted as $R_{j,j+1}$, while the distance between NNN atoms is given by $R_{j,j+2}=\sqrt{3}R_{j,j+1}$. We apply two global driving fields with Rabi frequencies $\Omega _{p_1}$ and $\Omega _{p_2}$ that correspond to detunings $\Delta_1$ and $\Delta_2$, respectively. The interaction between the NN atoms is denoted as $\mathcal{U}_{j,j+1}=\Delta_1$, and the interaction between NNN atoms is given by $\mathcal{U}_{j,j+2}=\Delta_2=\Delta_1/(\sqrt{3})^6$, as illustrated in Fig.~\ref{fig3}(b).
Additionally, for the three atoms of sublattice A, laser fields ($\Omega _1^{\alpha_1}$, $\Omega _2^{\beta_2}$), ($\Omega _2^{\alpha_2}$, $\Omega _3^{\beta_3}$), and ($\Omega _3^{\alpha_3}$, $\Omega _1^{\beta_1}$) are applied to the atoms at points $A_1$, $A_2$, and $A_3$, respectively. Similarly, laser fields are applied to lattice points $B_1$, $B_2$, and $B_3$ with pairs ($\Omega _1^{\beta_1}$, $\Omega _2^{\alpha_2}$), ($\Omega _2^{\beta_2}$, $\Omega _3^{\alpha_3}$), and ($\Omega _3^{\beta_3}$, $\Omega _1^{\alpha_1}$). All laser fields propagate along the quantization axis. This arrangement ensures that NN hopping occurs with the same phase $\alpha_j$ ($\beta_j$), while NNN hopping involves a phase difference of $\alpha_j-\beta_j$. As a result, a synthetic magnetic flux of ${\pi}/{2}$ is generated between triangles formed by each three-atom sublattice. In typical honeycomb tight-binding lattice models, a threefold rotational symmetry $\mathcal{C}3$ exists due to identical NN and identical NNN hopping strengths. However, this symmetry can be broken and an extended model is obtained by introducing unequal NN hopping strengths, such as $J_{j,j+1}^{(1)}<J_{j,j+1}^{(2)}$, as discussed in Ref.~\onlinecite{PhysRevB.104.L121108}.  To simulate this model, we can choose specific parameters as follows: $\Omega _1=\Omega _2=2\pi\times0.03$ MHz, $\Omega_3=2\pi \times0.0373$ MHz, $\Omega _{p_1}=2\pi \times0.95$ MHz, $\Omega _{p_2}=2\pi \times2.9$ MHz, $ \delta_1=-\delta_2=2\pi \times1$ MHz, $ \delta_3=2\pi \times1.8$ MHz, $\Delta_1=2\pi \times540$ MHz, and $\Delta_2=2\pi \times20$ MHz. Therefore, we obtain $J_{j,j+1}^{(1)}=2\pi \times1$ kHz, $J_{j,j+1}^{(2)}=2\pi \times0.48588$ kHz, and $J_{j,j+2}=2\pi \times0.47845$ kHz.

\section{Discussion}
\subsection{Experimental parameters}
\begin{figure}
    \centering
    \includegraphics[width=0.9\linewidth]{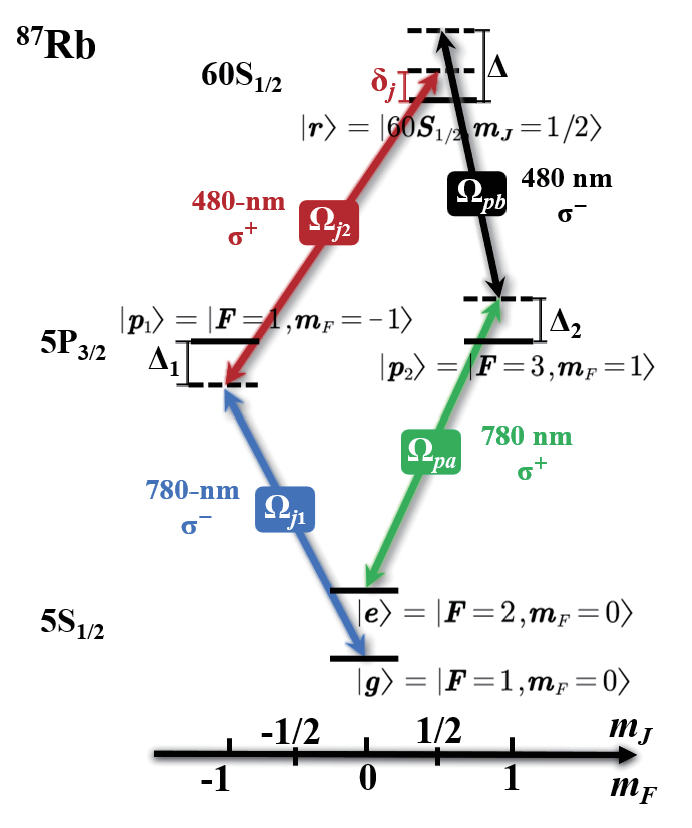}
    \caption{Level structure for the proposed atomic system in $^{87}$Rb and the laser-atom interaction model.}
\label{fig4}
\end{figure}

Now, we provide a concise discussion of the experimental feasibility of the proposed scheme. The transition from the ground states $|g\rangle$ or $|e\rangle$ to the Rydberg state $|r\rangle$ is facilitated through two-photon processes, as shown in Fig.~\ref{fig4}. Specifically, the ground state $|g\rangle$ undergoes coupling to the intermediate state $|5P_{3/2},\ F=1, m_F=-1\rangle$ using a local 780-nm $\sigma ^-$ polarized light. Subsequently, it is linked to $|r\rangle$ with another local 480-nm $\sigma ^+$ polarized light. On the other hand, the alternate ground state $|e\rangle$ experiences excitation through a 780-nm $\sigma ^+$ polarized light to reach the intermediate state $|p_2\rangle=|5P_{3/2},\ F=3, m_F=1\rangle$, and subsequently, it connects to the Rydberg state $|r\rangle$ with a 480-nm $\sigma^-$ polarized laser field. Given that the energy level difference between the two clock states $|g\rangle$ and $|e\rangle$ is $2\pi\times6.83~$GHz, this substantial energy gap, in conjunction with the polarization of the optical fields, effectively suppresses excitation through alternative pathways. Additionally, recent experiments demonstrate that atoms can be cooled close to their motional ground state.\cite{Tian:24} By utilizing counterpropagating lasers with distinct polarizations, the impact of the Doppler effect is further mitigated. Furthermore, because all laser fields propagate along the $z$ axis, the undesired relative phase between atoms caused by the laser can be neglected.

The configuration of each atom can be simplified to a three-level structure, as shown in Fig.~\ref{fig1}(b), by adiabatically eliminating the intermediate states $|p_{1(2)}\rangle$. This simplification is achieved under the conditions $(\Omega_{j1}\Omega_{j2})/(4\Delta_1) = \Omega_j$ and $(\Omega_{pa}\Omega_{pb})/(4\Delta_2) = \Omega_p$, which require $\Omega_{j1}/(2\pi) = \Omega_{j2}/(2\pi) \approx 40$--80 MHz, $\Omega_{pa}/(2\pi) = \Omega_{pb}/(2\pi) \approx 179.5$ MHz, and $\Delta_1/(2\pi) = \Delta_2/(2\pi) \approx 2$ GHz.
Specifically, $\Omega_{j1}/(2\pi) \approx 40$--80 MHz corresponds to a beam power $P_{j1}$ of 12.31--49.26 nW and a waist of 1.5 $\mu$m, while $\Omega_{j2}/(2\pi) \approx$ 40--80 MHz corresponds to a beam power $P_{j2}$ of 9.42--37.71 mW and a waist of 1.5 $\mu$m. Similarly, $\Omega_{pa}/(2\pi) \approx 179.5$ MHz corresponds to a beam power $P_{pa}$ of 11.48 $\mu$W and a waist of 10 $\mu$m, and $\Omega_{pb}/(2\pi) \approx 179.5$ MHz corresponds to a beam power $P_{pb}$ of 899.67 mW and a waist of 4 $\mu$m.

\subsection{Undesired interatomic distance}
\begin{figure}
    \centering
    \includegraphics[width=1\linewidth]{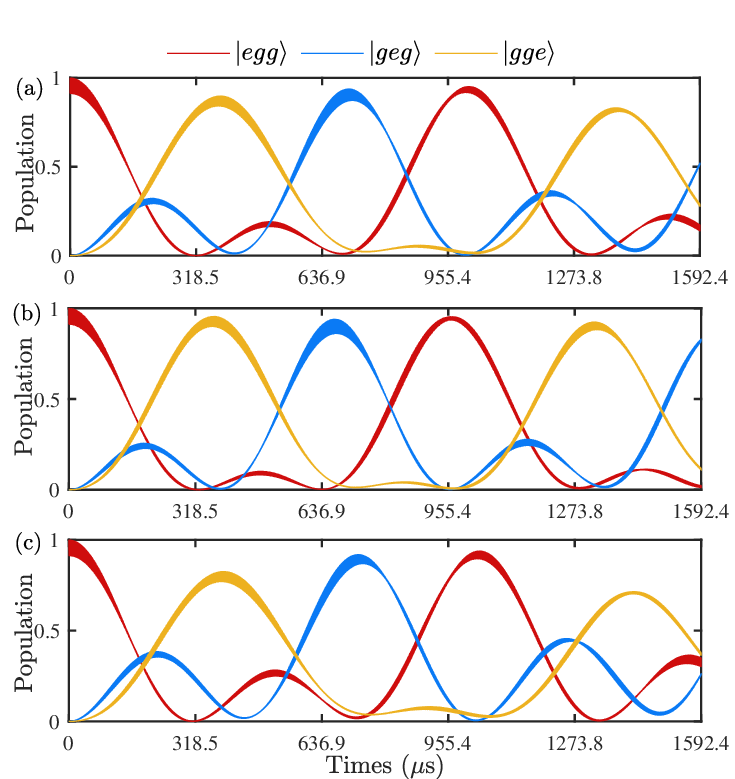}
    \caption{Population oscillations of three-atom states $|gee \rangle$, $|ege \rangle$ and $|eeg \rangle$. (a) $r_1\approx4.3447$ $\mu {\rm m}$ with $\mathcal{U}_{j,j+2}=2\pi \times20$ MHz $=\Delta_2$. (b) $r_2\approx4.3647$ $\mu {\rm m}$ with $\mathcal{U}_{j,j+2}=2\pi \times19.456$ MHz. (c) $r_3\approx4.3247$ $\mu {\rm m}$ with $\mathcal{U}_{j,j+2}=2\pi \times20.561$ MHz. The parameters are $(\Omega _1,\Omega _2,\Omega _3,\Omega _p)/(2\pi)=(0.03,0.03,0.0373,2.9)$ MHz, $( \delta _1,\delta _2,\delta_3,\Delta_2)/ (2\pi)=(1,-1,1.8,20)$ MHz.}
\label{fig5}
\end{figure}

In the preceding sections, we exclusively focused on the system characterized by the desired interatomic distance, where the Rydberg interaction strength equates to the corresponding blue-detuning parameter of the two-photon transition. However, it is essential to acknowledge that atoms exhibit positional fluctuations attributable to the finite temperature of traps and the repulsion induced by the vdW force. The incorporation of these position fluctuations for simulating multiple random quantum state evolution trajectories and subsequent averaging undoubtedly entails a significantly time-intensive process. To qualitatively explore the impact of imperfect atomic spacing, we adopt a fixed but undesired interatomic distance uniformly altered within the equilateral triangle structure depicted in Fig.~\ref{fig3}(a), composed of $A_1-A_2-A_3$ or $B_1-B_2-B_3$. 
In Fig.~\ref{fig5}(a), we plot the population oscillations of the three-atom states $|gee \rangle$, $|ege \rangle$, and $|eeg \rangle$ at the desired interatomic distance $r_1\approx4.3447$ $\mu {\rm m}$, satisfying $\Delta_2=\mathcal{U}_{j,j+2}=2\pi \times20$ MHz. Figs.~\ref{fig5}(b) and~\ref{fig5}(c) illustrate the corresponding outcomes for $r_1\pm0.02$ $\mu {\rm m}$, respectively. Evidently, within the considered range of atomic spacing, the periods of quantum state population evolution remain essentially consistent across all three cases, indicating no discernible dephasing effects. Nonetheless, this finding also underscores the necessity for stricter temperature control within the trap.

\subsection{Coherence time of the system}
\begin{figure}
    \centering
    \includegraphics[width=1\linewidth]{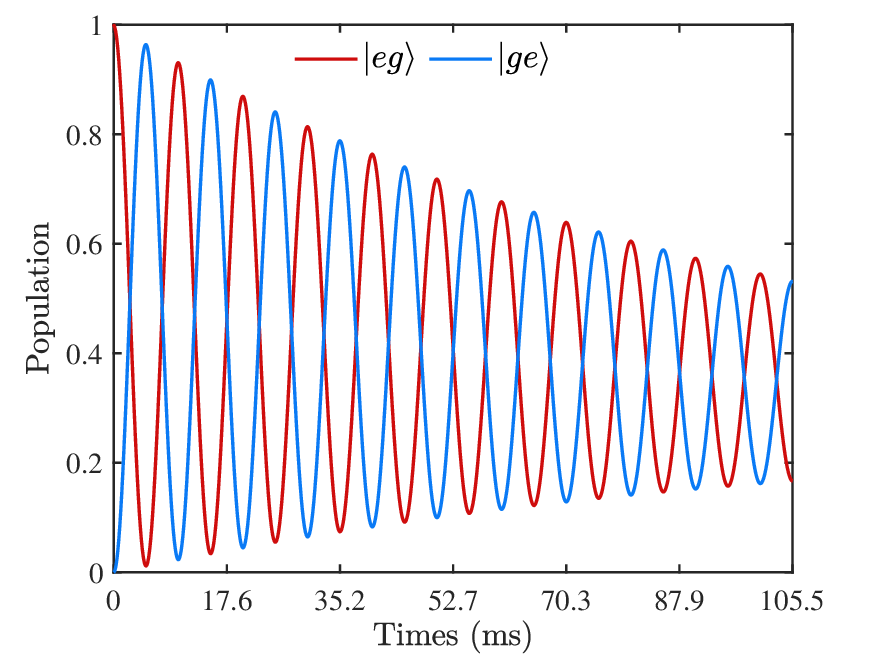}
    \caption{Rabi oscillations of two atoms states $|eg \rangle$ and $|ge \rangle$ dominated by the master equation of Eq.~(\ref{C2}).}
\label{fig6}
\end{figure}

The Rabi oscillations between any pair of two atoms can be used to directly measure the coherence time of our model. As an example,  we consider the dynamics of atoms $A_1$ and $A_2$ in Fig.~\ref{fig3}(a). The effective flip-flop coupling strength between them is $2\pi \times0.47845$~kHz, which corresponds to the full Hamiltonian of Eq.~(\ref{Hamiltonian_1}) with parameters $(\Omega_2,\Omega _p,\delta_2,\Delta)/(2\pi)=(0.0373,2.9,1,20)$ MHz, as given by Sec.~\ref{SecIII}.
Specifically, 
the evolution of the system is governed by the Markovian master equation
\begin{equation}\label{C2}
\partial _t\rho =-i\left[ H_I^{(2)},\rho \right]+\sum_{j=1}^6{L_j\rho L_j^{\dag}-\frac{1}{2}\left\{ L_j^{\dag}L_j,\rho \right\}},
\end{equation}
where the Hamiltonian of the two atoms reads 
\begin{eqnarray}
H_I^{(2)}&=&\Omega _2^{\beta_2}e^{-i\delta _2t}|r_1 \rangle\langle g_1|+\Omega _2^{\alpha_2}e^{-i\delta _2t} | r_2 \rangle\langle g_2|\nonumber\\&&
+\Omega _p \sum_{j=2}^3 |r_j \rangle\langle e_j|e^{-i\Delta t}+{\rm H.c.}+\mathcal{U}_{1,2}|r_1r_2 \rangle\langle r_1r_2|,\nonumber\\&&
\end{eqnarray}
and 
$L_1=\sqrt{\gamma_1/2}|g_1\rangle\langle r_1|$, $L_2=\sqrt{\gamma_1/2}|e_1\rangle\langle r_1|$, $L_3=\sqrt{\gamma_1/2}|g_2\rangle\langle r_2|$, and $L_4=\sqrt{\gamma_1/2}|e_2\rangle\langle r_2|$ with $\gamma_1/(2\pi)=6.92$ kHz  characterize the spontaneous emission of the Rydberg states to the computational basis states. Additionally, we have included the dephasing operators $L_5=\sqrt{\gamma_2}|r_1\rangle\langle r_1|$ and $L_6=\sqrt{\gamma_2}|r_2\rangle\langle r_2|$, where $\gamma_2/(2\pi)=20$ kHz, to account for laser phase noise and other noise sources,\cite{PhysRevA.99.043404,PhysRevX.14.011025} for the sake of simplicity. We depict the population oscillations of states $|eg\rangle$ and $|ge\rangle$ versus the evolution time in Fig.~\ref{fig6}.  It can be observed that the $1/e$ coherence time of the system is 105.5 ms, which corresponds to 11.5 Rabi oscillations.

\section{summary}
In summary, by leveraging multi-color driving fields and an unconventional Rydberg pumping mechanism, we have developed a theoretical framework for achieving chiral motion of excitation in a three-atom system. This scheme offers highly adjustable parameters and successfully eliminates the previous reliance on modulation periods, as seen in conventional Floquet methods. Consequently, its applicability can be readily extended to multi-particle systems, such as the 2D triangular array and the honeycomb lattice. We also discussed the details of our proposed scheme, including the specific experimental setup and corresponding parameters. We found that imperfect atomic spacing does not significantly affect the period of population evolution, with no discernible dephasing effects. Additionally, we examined the coherence time of our model by considering the Rabi oscillations between any pair of two atoms. Compared to previously proposed schemes involving Rydberg atoms, we encode quantum information in the ground-state manifolds of neutral atoms and employ dispersion coupling between lasers and atoms, effectively minimizing spontaneous emission and mitigating the adverse effects of Rydberg state excitation. With these advancements, we anticipate that our proposed scheme can be experimentally implemented in the near future.

This work was supported by the National Natural Science Foundation of China (NSFC) under Grant No. 12174048, the Fundamental Research Funds for the Central Universities
under Grant No. 3122023QD26, and the National Research Foundation Singapore (NRF2021-QEP2-02-P01), A*STAR Career Development Award (C210112010), and A*STAR (C230917003, C230917007).

\vspace{12pt}
\section*{AUTHOR DECLARATIONS}
\subsection*{Conflict of Interest}
The authors have no conflicts to disclose.

\vspace{12pt}
\subsection*{Author Contributions}
{\bf Hao-Yuan Tang}: Data curation (equal); 
 Writing-original draft (lead). {\bf Xiao-Xuan Li}: Data curation (equal). {\bf Jia-Bin You}: Writing-review \&
editing (equal);  Supervision (equal); Funding acquisition  (equal). {\bf Xiao-Qiang Shao}: Conceptualization (lead); 
 Writing-review \&
editing (equal); Supervision (equal); Funding acquisition  (equal). 

\vspace{12pt}
\section*{DATA AVAILABILITY}
The data that support the findings of this study are available from the corresponding authors upon reasonable request.

\appendix

\section{The details of Hamiltonian (3)}\label{Ap3}
\begin{figure}
    \centering
    \includegraphics[width=1\linewidth]{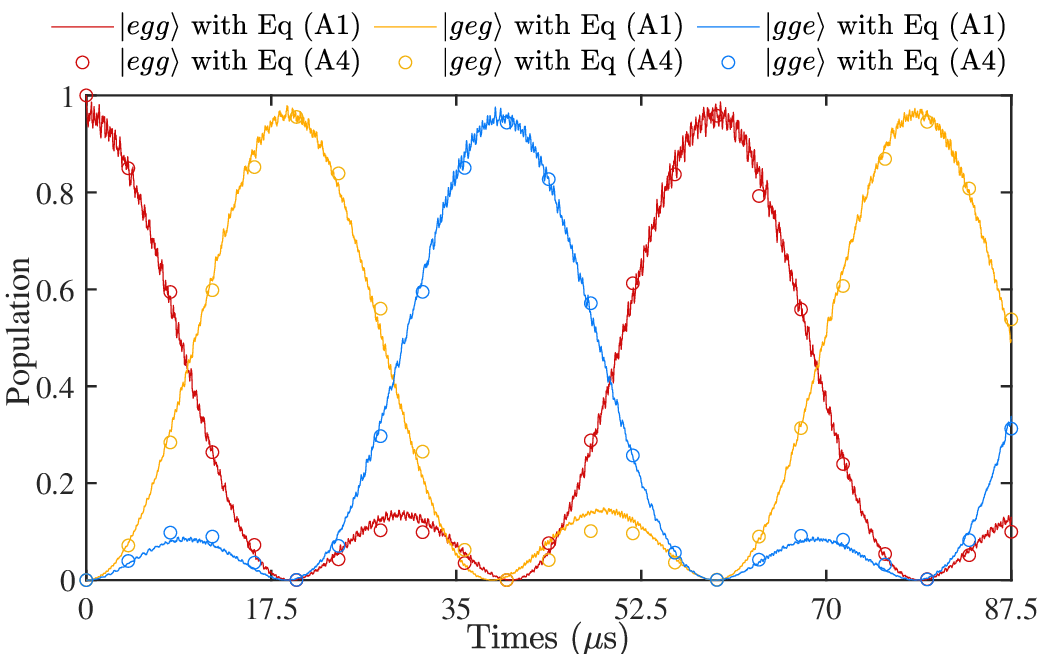}
    \caption{The chiral motion of atomic excitation among the system governed by the full Hamiltonian~(\ref{A1}) and Eq.~(\ref{A4}), respectively. The parameters are $(\Omega _1,\Omega _2,\Omega _3,\Omega _p)/(2\pi)=(0.2,0.2,0.4,4)$ MHz, $(\alpha_1,\alpha_{j\ne 1},\beta_j)=(\pi/{2},0,0)$, and $( \delta _1,\delta _2,\delta_3,\Delta)/ (2\pi)=(-4,4,8,200)$ MHz.}
\label{fig7}
\end{figure}

The full Hamiltonian of the equilateral triangle structure shown in Fig.~\ref{fig1}(a) reads
\begin{eqnarray}\label{A1}
H_I&=&(\Omega _1^{\alpha_1}e^{-i\delta _1t}+\Omega _2^{\beta_2}e^{-i\delta _2t})| r_1 \rangle\langle g_1|+(\Omega _2^{\alpha_2}e^{-i\delta _2t} \nonumber\\&&
+\Omega _3^{\beta_3}e^{-i\delta _3t})| r_2 \rangle\langle g_2|+(\Omega _3^{\alpha_3}e^{-i\delta _3t}+\Omega _1^{\beta_1}e^{-i\delta _1t})| r_3 \rangle\langle g_3|) \nonumber\\&&
+\Omega _p \sum_{j=1}^3 |r_j \rangle\langle e_j|e^{-i\Delta t}+{\rm H.c.}+\frac{1}{2}\sum_{i\neq j}\mathcal{U}_{i,j} |r_ir_{j} \rangle\langle r_ir_{j}|.\nonumber\\&&
\end{eqnarray}
Considering a rotating frame
with respect to $U=\exp[-i\sum_{i\neq j}\mathcal{U}_{i,j}/2|r_ir_{j} \rangle\langle r_ir_{j}|t]$, and $\mathcal{U}_{i,j}=\Delta$. The Hamiltonian satisfies $H_{I}^{'}=i\dot{U}^{\dag}U+U^{\dag}H_IU$, then the Eq.~(\ref{A1}) can be reformulated as:

\begin{eqnarray}\label{A2}
H'_{I}&=& \sum_{j=1}^3  \Omega _{j}^{\alpha_j}\sum_{k,k^{'}=g,e}{\sigma}_{j}^{rg}[ P_{j+1}^{k}P_{j+2}^{k^{'}}e^{-i\delta_jt}+( P_{j+1}^{r}P_{j+2}^{k} \nonumber\\&&
+P_{j+1}^{k^{'}}P_{j+2}^{r} ) e^{i\phi_jt} ]+\Omega _{j+1}^{\beta_{j+1}}\sum_{k,k^{'}=g,e}{\sigma}_{j}^{rg}[ P_{j+1}^{k}P_{j+2}^{k^{'}}e^{-i\delta _{j+1}t} \nonumber\\&&
+( P_{j+1}^{r}P_{j+2}^{k}+P_{j+1}^{k^{'}}P_{j+2}^{r} ) e^{i\phi_{j+1}t} ]+(\Omega _{j}^{\alpha _j}\sigma _{j}^{rg}P_{j+1}^{r}P_{j+2}^{r}\nonumber\\&&
+\Omega _{j}^{\beta _j}\sigma _{j+2}^{rg}P_{j+1}^{r}P_{j}^{r})e^{i\theta_j t}+\Omega_p[\sum_{k=g,e}P_{j}^{r}(\sigma _{j+1}^{re}P_{j+2}^{k} \nonumber\\&&
+P_{j+1}^{k}\sigma _{j+2}^{re})+\sum_{k,k^{'}=g,e}\sigma _{j}^{re}P_{j+1}^{k}P_{j+2}^{k^{'}}e^{-i\Delta t}\nonumber\\&&
+\sigma _{j}^{re}P_{j+1}^{r}P_{j+2}^{r}e^{i\Delta t} ]+{\rm H.c.},
\end{eqnarray}
where $\phi_i=\Delta-\delta _i$ and $\theta_i=2\Delta-\delta _i$. The high-frequency oscillation term can be disregarded due to the limit of large detuning $\Delta \gg \{ \Omega _p, \Omega _j, \delta _j \}$, then the Eq.~(\ref{A2}) is reduced to
\begin{eqnarray}\label{A3}
H'_{I}&=& \sum_{j=1}^3  \Omega _{j}^{\alpha_j}\sum_{k,k^{'}=g,e}{\sigma}_{j}^{rg} P_{j+1}^{k}P_{j+2}^{k^{'}}e^{-i\delta_jt}\nonumber\\&&
+\Omega _{j+1}^{\beta_{j+1}}\sum_{k,k^{'}=g,e}{\sigma}_{j}^{rg} P_{j+1}^{k}P_{j+2}^{k^{'}}e^{-i\delta _{j+1}t}+\Omega _{j}^{\alpha _j}\sigma _{j}^{rg}P_{j+1}^{r}P_{j+2}^{r} \nonumber\\&&
+\Omega_p\sum_{k=g,e}P_{j}^{r}(\sigma _{j+1}^{re}P_{j+2}^{k}+P_{j+1}^{k}\sigma _{j+2}^{re})+{\rm H.c.}.
\end{eqnarray}
The dynamics of the system we are considering are defined within three different subspaces $\{|egr\rangle, |rgr\rangle, |rge\rangle\}$, $\{|erg\rangle, |rrg\rangle, |reg\rangle\}$, and $\{|gre\rangle, |grr\rangle, |ger\rangle\}$.  Thus, Eq.~(\ref{A3}) can be further rewritten as
\begin{eqnarray}\label{A4}
H'_{I}&=&\sum_{j=1}^3  \Omega _{j}^{\alpha_j}{\sigma}_{j}^{rg}(P_{j+1}^{e}P_{j+2}^{g}+P_{j+1}^{g}P_{j+2}^{e})e^{-i\delta_jt}\nonumber\\&&
+\Omega _{j+1}^{\beta_{j+1}}{\sigma}_{j}^{rg}(P_{j+1}^{e}P_{j+2}^{g}+P_{j+1}^{g}P_{j+2}^{e})e^{-i\delta_{j+1}t} \nonumber\\&&
+\Omega_p \sigma _{j}^{re}(P_{j+1}^{g}P_{j+2}^{r}+P_{j+1}^{r}P_{j+2}^{g})+{\rm H.c.}.
\end{eqnarray}
To verify the accuracy of our effective model, we compared the dynamics of the system governed by the full Hamiltonian~(\ref{A1}) with those described by Eq.~(\ref{A4}), as shown in Fig.~\ref{fig7}. We found that these two cases match very well.

\vspace{12pt}
\section*{REFERENCES}
\bibliography{aipsamp.bbl}

\end{document}